\documentclass{aa}
\usepackage{graphicx}
\usepackage{txfonts}

\begin{document}

\title{$\Lambda$CDM Halo Density Profiles: where do actual halos converge to NFW ones?}
\author{Gianfranco Gentile\inst{1,2}, Chiara Tonini\inst{1} and Paolo
  Salucci\inst{1}}
\institute{SISSA, via Beirut 4, 34014 Trieste, Italy\\
              \email{ggentile@unm.edu}
         \and
             University of New Mexico, Department of Physics and Astronomy, 
             800 Yale Blvd NE, Albuquerque, NM 87131, USA
             }

\date{Accepted. Received}

\abstract
{}
{We present an analysis of 37 high-quality extended rotation curves
that highlights the existence of a new discrepancy 
(or a new aspect of an old discrepancy) between the
density profiles predicted by the $\Lambda$ Cold Dark Matter ($\Lambda$CDM) 
theory and the actual distribution of dark matter in galaxies.}
{We compare the predicted face-value density vs. enclosed mass relationship,
at large distances, to the observational data at the last measured
radii of the rotation curves and in two whole rotation curves of high
quality. A further analysis is performed
by studying a relation, inbuilt in $\Lambda$CDM,
that links at radius $R$, the enclosed halo mass 
$M_{\rm NFW}(R)$ and its density $\rho(R)$
in a way that is independent of the mass of the virialised object.}
{We find that the predicted density vs. enclosed mass relationship
has a systematic offset with respect to the observational data.
In test case extended rotation curves,
at their last measured point,
the predicted NFW densities are up to a factor 3 lower 
than those derived from the kinematics. Moreover, the abovementioned
relation, inbuilt in $\Lambda$CDM, does not hold for the objects of our sample.
Such a new outer discrepancy is different and maybe complementary with
respect to the core/cusp issue, for which the NFW 
densities turn out to be higher 
than those observed 
and it seems to imply a global mass rearrangement of a pristine
NFW-$\Lambda$CDM halo.}
{}


\titlerunning{Outer $\Lambda$CDM Halo Density Profiles}
\maketitle

\section{Introduction}
\label{intro}

The study of disk galaxies' rotation curves has been one of the
most successful tools for investigating the dark matter phenomenon
in galaxies. Recently, the attention has been focused on the distribution
of dark matter in galaxies, as inferred from the rotation curves, in particular 
the comparison between
the predictions of standard $\Lambda$ Cold Dark Matter ($\Lambda$CDM) theory of
structure formation and observations (e.g., Salucci \& Burkert 2000,
Borriello \& Salucci 2001,
de Blok \& Bosma 2002, Weldrake et al. 2003, Swaters et al. 2003,
Simon et al. 2005, Gentile et al. 2004, 2005, 2006).

More specifically, the $\Lambda$CDM theory predicts that dark matter halos have a specific
density distribution that follows the well-known NFW (Navarro, Frenk and White, 1996) profile:

\begin{equation}
\rho_{\rm NFW}(R) = \frac{\rho_{\rm s}}{(R/r_{\rm s})(1+R/r_{\rm s})^2}
\label{eq:rho_nfw}
\end{equation}

where $r_{\rm s}$ and $\rho_{\rm s}$ are the characteristic radius and density
of the distribution. The latter is given by: 

\begin{equation}
\label{rhosnfw}
\rho_{\rm s} = \frac{\Delta}{3} \frac{c^3}{{\rm ln}(1+c)-\frac{c}{1+c}} \rho_{c},
\label{eq:wechsler2}
\end{equation} 

where $\rho_{\rm c}$ is the critical density of the Universe
and $\Delta$ is the virial overdensity (see Bryan \& Norman 1998).

$r_{\rm s}$ and $\rho_{\rm s}$ are related to each other (e.g. Wechsler et al. 2002),
so eq. 1 is rather a one-parameter family of profiles, where the following relations
link the virial mass $M_{\rm vir}$ to the concentration parameter
$c$~(=$r_{\rm vir}/r_{\rm s}$, where $r_{\rm vir}$ is the virial radius), 
$r_{\rm s}$ and $\rho_{\rm s}$, at redshift $z=0$
and adapting the relations (similarly to Gnedin et al. 2006 and Dutton et al. 2006)
to the cosmological parameters from the WMAP third year results (Spergel et
al. 2006):

\begin{equation}
\label{cmvir}
c \simeq 13.6 \left( \frac{M_{\rm vir}}{10^{11} {\rm M_{\odot}}} \right)^{-0.13},~~ 
r_{\rm s} \simeq 8.8 \left( \frac{M_{\rm vir}}{10^{11} {\rm M_{\odot}}} \right)^{0.46}~{\rm kpc}
\label{eq:wechsler1}
\end{equation} 

The virial radius $r_{\rm vir}$ can then be derive from 
$r_{\rm vir}=c~r_{\rm s}$.
In $\Lambda$CDM, at least in a statistical sense, 
once the radius and the mass
at that radius are fixed, the parameter describing the mass distribution
(usually the virial mass $M_{\rm vir}$) is also known.
On the other hand, a quite remarkable number of
observations show that NFW profiles, displaying an inner ``cusp'',
are inconsistent with data. In fact, the latter indicate profiles with a different
characteristic, a central density ``core'', i.e. a region where the dark matter
density remains approximately constant. 
A number of studies cast doubts on the reliability of the mass modelling
procedure and the data analysis (van den Bosch et al. 2000, Swaters et
al. 2003, Hayashi \& Navarro 2006, Valenzuela et al. 2007)
of some galaxies. 
The debate is still on,
despite the fact that most concerns are now overcome
(Gentile et al. 2004, 2005, 2006, de Blok, Bosma \& McGaugh 2003, de Blok
2005). 

The simplest example of a cored halo profile
is the pseudo-isothermal (PI, van Albada et al. 1985): 
\begin{equation}
\rho_{\rm PI}(R) = \frac{\rho_0}{1+R^2/R^2_{\rm C}}
\label{eq:rho_iso}
\end{equation}

where $R_{\rm C}$ is the core radius and $\rho_0$ is the central density, that
results about one order
of magnitude lower than $\Lambda$CDM predictions (Donato et al. 2004, Gentile et al. 2004,
2005 and references therein).

   \begin{figure*}
   \centering
   \includegraphics[width=9.5cm]{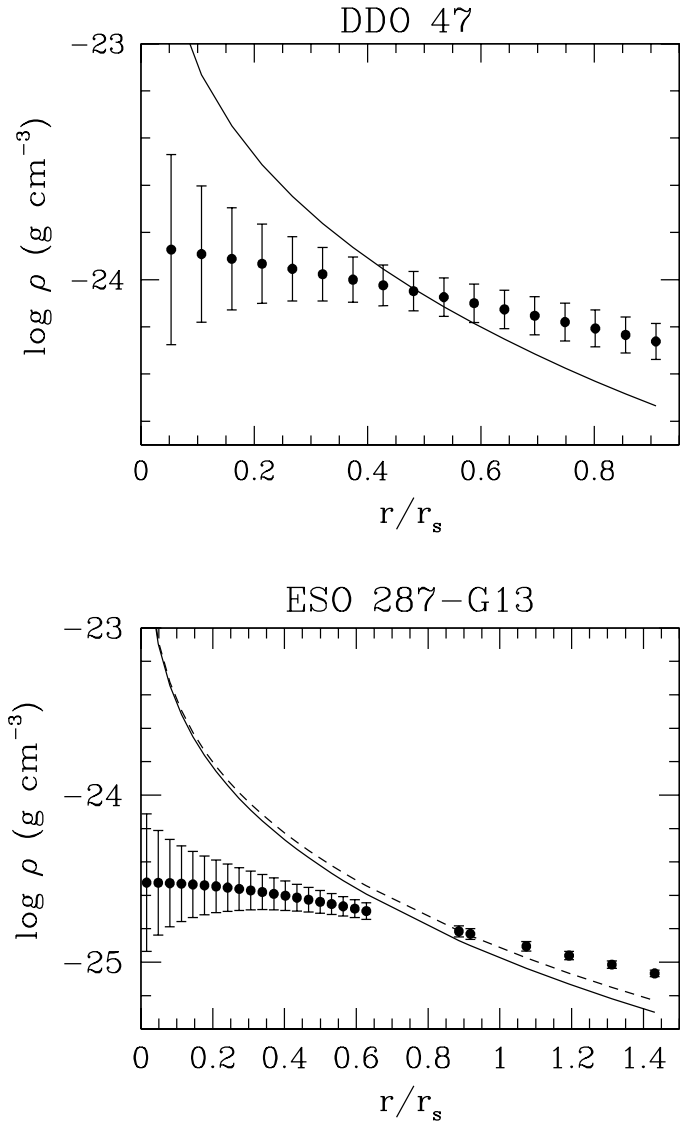}  
   \caption{DM density distributions;
filled circles represent the DM density inferred by the 
rotation curves.
Solid lines corresponds to the NFW solution with $M_{\rm NFW}(R_f)=M_{\rm PI}(R_{\rm f})$,
that give, for DDO 47, ($r_{\rm s}$, $M/L_{\rm B}$; $R_{\rm D}$)=(5.5 kpc, 0.5; 0.5 kpc),
and for ESO 287-G13, ($r_{\rm s}$, $M/L_{\rm I}$; $R_{\rm D}$)=(17.4 kpc, 1.8; 3.3 kpc).,
The dashed line corresponds to the ``best fit'' NFW solution, ($r_{\rm s}$,
$M/L_{\rm I}$)=(19.4 kpc, 0.7). See text for the (conservative) estimate of the 
uncertainties.
}
    \label{densprof}
    \end{figure*}

However, the above works did focus on the comparison between $\Lambda$CDM and
observations in the {\it inner} parts of the galaxies, while it is of extreme 
importance to investigate also the {\it outer} density distribution, 
where $\Lambda$CDM halos have an equally
strong feature: the transition between a $\rho \propto R^{-1}$ to a $\rho
\propto R^{-3}$ regime.
In this paper, outer and inner regions  have a 
baryonic perspective. The former refers to a region inside $1 - 2$ 
stellar disk scale lengths $R_{\rm D}$, the latter to that extended out to 
$\gtrsim 3 - 5 R_{\rm D}$
and characterised by the HI disk. Both regions, 
however, must be considered inner regions with respect to the DM 
distributions.
In other words, is the discrepancy between theoretical predictions
and data present only in the inner parts of a galaxy, and therefore the profiles converge
to the NFW one at larger radii, or is there a more global 
discrepancy extending beyond the NFW halo $R^{-1}$ regime,
as suggested also by McGaugh et al. (2006)? 

Note that the rotation curves at large radii, i.e. for $R>> R_{\rm D}$, 
provide a very good measure of 
densities and of enclosed dark masses. In fact, at these distances, the contribution to the
gravitational potential due to gas and stars is very small and the uncertainties
on the accurateness/uniqueness  of the dark-to-luminous
mass decomposition play a very minor role: it is the rotation curve itself  that essentially measures
the physical quantities $
M_{\rm h}(R)\simeq  G^{-1} V^2(R) R $ and $\rho_{\rm h} (R)\simeq 1/(4 \pi R^2) dM_{\rm h}(R)/dR $. 
Conversely, in the inner regions 
($R < R_{\rm D}$), usually, a complex dark-luminous mass modelling is needed to decompose the circular velocity 
into the (possibly equally important) dark and luminous contributions (see e.g. 
Barnes, Sellwood \& Kosowsky 2004). 

\section{Samples and methods of investigation}

In the present paper we investigate the outer dark matter distribution
mostly resorting to two samples, heterogeneous in mass, covering about 3
orders of magnitude: 1) the sample of high-quality
rotation curves selected by Donato et al. (2004), discarding the 4 galaxies with the 
smallest extension relative to the disk exponential
scale length $R_{\rm D}$ 
and adding the galaxies: DDO 47 (Salucci, Walter and
Borriello 2003, Gentile et al. 2005) and 
ESO 287-G13 (Gentile et al. 2004); 2) a sample with rotation curves 
selected from the literature fulfilling the
requirements of a) reaching out to at least 30 kpc, or out to 6 disk scale
lengths, or b) with the final velocity being above $250$ km s$^{-1}$. 
In this way, the
curves were sufficiently extended to map a region of the halo density
profile not affected by the central slope, nor massively affected by the
presence of the disk. It results that we investigate regions of halos
corresponding to galactocentric distances out to $5\%$ up to about $35\%$ (the case
of NGC 9133) of the virial
radius: these zones are beyond the influence of the cusp, and still well
into the central part of the dark matter halo.
We selected only rotation 
curves that were regular out to the last measured radius $R_{\rm f}$ (whose average value
is about 24 kpc), obtaining a sample of 37
galaxies. In Table 1 the selected galaxies of sample 2 are listed, together with the 
corresponding references. 

For a number of galaxies the mass decomposition between the
luminous and dark component of the velocity was obtained from the
literature; we marked them 
with an asterisk in Table 1. For the others, 
we used the method of
Persic and Salucci (1990), see Appendix A. 

 \begin{table}
\centering
\caption{Selected galaxies with their references. The asterisk indicates that
  the original work provided also the dark-luminous decomposition of the rotation curve.}
\begin{tabular}{l|l}
\hline
Galaxy & Reference\\
\hline
NGC $289^*$ & Walsh et al., 1997\\
NGC $1068$ & Sofue et al., 1999\\
NGC $1097$ & Sofue et al., 1999\\
NGC $1232^*$ & van Zee \& Bryant, 1999\\
NGC $3198^*$ & Blais-Ouellette et al., 2001 \\
NGC $3726$ & Verheijen \& Sancisi, 2001\\
NGC $4123^*$ & Weiner et al., 2001\\
NGC $5055$ & Sofue et al., 1999\\
NGC $5236$ & Sofue et al., 1999\\
UGC $5253$ & Noordermeer et al., 2004\\
NGC $5985$ & Blais-Ouellette et al., 2004 \\
NGC $6946^*$ & Carignan et al., 1990 \\
NGC $7331^*$ & Bottema, 1999\\
UGC $9133^*$ & Noordermeer et al., 2004\\
\hline   
\end{tabular}
\end{table}

The rotation curves of sample 1
have been selected and successfully used (Donato et al. 2004) in order to investigate the core
radius issue, i.e. an issue that needs 
more accurate data than the issue we want to tackle here. 
Since the resulting halo rotation curves of the galaxies of the sample are very well fitted 
by the PI profile, we will take it 
as representative of the dark matter haloes around galaxies. 

From our sample 1 we select DDO 47 and ESO 287-G13, as the best 
examples to show individually the outer NFW $\Lambda$CDM-data discrepancy.
Then, we use the combined samples 1-2 to compare, 
at the outermost radii, the NFW $\Lambda$CDM predictions,
obtained by a newly discovered structural $\Lambda$CDM 
relation, with the values of the mass, radius and density 
of the DM halos around galaxies.

\section{The new dark matter density discrepancy from the extended rotation curves
of ESO 287-G13 and DDO 47}

We analyse in detail the density profiles of the DM halos for the two best cases in our sample: 
DDO 47 (Salucci, Walter and Borriello 2003, Gentile et al. 2005) and ESO 287-G13 (Gentile et al. 2004).
Note that in the  present study, differently from the original papers, 
the aim is to show the existence of an
outer observations/theory discrepancy, rather than to investigate the inner cusp/core issue: 
therefore we will analyse the data in a different but proper way.
For consistency with the detailed analysis made in the original papers,
in this Section we used the relation between $c$ and $M_{\rm vir}$
given by Bullock et al. (2001) and Wechsler et al. (2002) instead of
eq. \ref{eq:wechsler1}.

\subsection{Actual dark matter halos}

We follow Salucci et al. (2003) and Gentile et al. (2004), 
the density distribution $\rho_{\rm h}(R)$ of the dark 
matter halo in these galaxies is given by:
$\rho_{\rm h}(R)=\rho_{\rm PI}(R)$ 
(see eq. 4). Notice that in both objects we have
also considered a Burkert halo (Burkert 1995), and found results coincident with
the PI halo. 

\subsection{NFW halos}

Let us assume that: 

\begin{equation}
M_{\rm NFW}(R_{\rm f})=M_{\rm PI}(R_{\rm f})
\label{masseq}
\end{equation}

where $M_{\rm PI}(R_{\rm f})$ is the mass inside $R_{\rm f}$
we derive from the above mass model. 
Eq. \ref{masseq} allows
to derive the values for $c$ and $M_{\rm vir}$: 18.4 and 6 $\times~10^{10}$ 
$M_{\odot}$ for DDO 47, and
13.3 and 7 $\times~10^{11}$ $M_{\odot}$ for ESO 287-G13. 
Let us anticipate that
this assumption is very conservative;
in fact, if the mass equality expressed by 
eq. \ref{masseq} occurs at radii smaller than $R_{\rm f}$ 
the result we claim here will 
be even more prominent; on the other hand,
if the mass equality occurs at radii larger than $R_{\rm f}$, a 
discrepancy from $R=0$ out to $R=R_{\rm f}$ will be set by definition. 
This assumption has also the desirable
by-product, in the NFW framework,
of implying reasonable values for the stellar mass-to-light ($M/L$)
ratios. 

Given the importance of a careful analysis, we have also considered
a different implementation of the NFW halo + disk + gas
mass modelling: in ESO 287-G13 we have fitted the rotation curve 
with this mass model by
leaving $M/L$ and
$M_{\rm vir}$ as free parameters;
this latter procedure, shown in Fig. \ref{densprof}, yields very similar results with respect
to those obtained by means of eq. \ref{masseq}.

\subsection{Test for ESO 287-G13 and DDO 47}

In Fig. \ref{densprof} we compare the actual dark matter density 
$\rho_{\rm PI}(R)$ and $\rho_{\rm NFW}(R)$
at the radii where the rotation curves were measured.
A conservative estimate of the uncertainty on the 
density was derived 
from the mass modelling uncertainties. In fact 20\% uncertainties in 
$\rho_0$ and $R_{\rm C}$ give an average uncertainty on 
the density of about 25\%. 
A rigorous derivation goes beyond the scope of this paper.
Then, it is clear that the density
is not determined with the same accuracy at each radius. 
In fact dark matter is not the dominant kinematic term in the
inner parts and therefore the density uncertainty is larger.

\begin{figure}
\centering
\includegraphics[width=8.2cm]{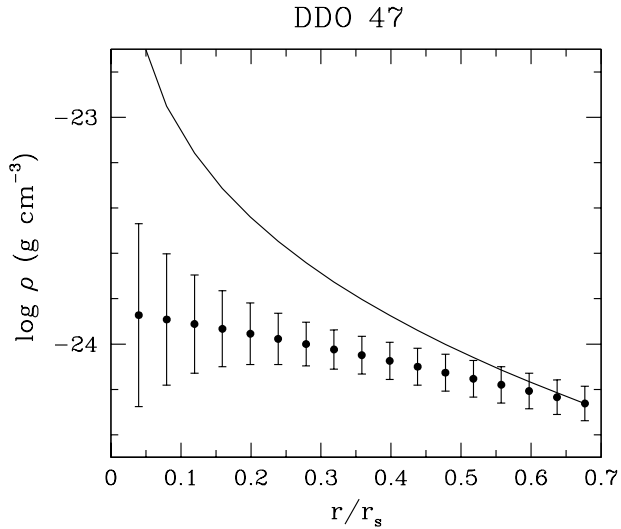}    
\caption{
Density profiles of DDO 47 assuming $\rho_{\rm NFW}(R_{\rm f})=\rho(R_{\rm f})$.
See Fig. \ref{densprof} for the explanation of the symbols.
}
\label{flattening}
\end{figure}

We see the well-known (though debated in some case) cusp-core discrepancy at $R \rightarrow 0$,
but at larger radii we now realise that the difference in densities changes
sign and that the ``inner'' discrepancy is now reversed:
in both analyses, 
the NFW halo densities, from a
certain radius onward,  
are lower 
than the dark matter halo density. 
Note that such a discrepancy was also present in some of the analysis made
in previous investigations
(e.g. Blais-Ouellette et al. 2001, Borriello \& Salucci 2001, 
de Blok \& Bosma 2002),
but it was not claimed explicitely, neither it was investigated whether
some combination of $c$ and $M_{\rm vir}$
had made possible for $\rho_{\rm NFW}$ to converge to the actual density
{\it inside} the region mapped by the rotation curve data.
In the present paper we instead claim
that the situation can
be described by one of these two possibilities: 1) $\rho_{\rm NFW}(R)$
at any radius is systematically higher than the estimated
density
(so it will disagree in the core region, but will be compatible with the
outermost density, see Fig. \ref{flattening}) 
and the data vs. prediction discrepancy will extend
to several disk scale lengths; 2) the discrepancy is ``bivariate'': 
actual halos (in comparison to predicted halos) have a density deficit in the inner
regions but an excess in the outer ones.
We consider the second possibility as more likely since 
it implies stellar $M/L$ ratios compatible with the predictions
of stellar population synthesis models (Bell et al. 2003, Gentile et
al. 2004). For instance, in the case of DDO 47, imposing $\rho_{\rm NFW}(R)=
\rho(R)$ leads to a best-fit $M/L$ ratio of 0; on the other hand, with
the second possibility one gets a B-band $M/L=0.7$, within the range 
$0.5 - 0.8$ arising from the observed $B-V$ colour and the predictions
of stellar population synthesis models (Bell et al. 2003).

Let us stress that the
availability of data at large distances is a crucial point: in these galaxies we just reach
the radius where the NFW density decreases in an appreciable way.
Compared to the discrepancy in the inner regions of galaxies, extremely
evident and as wide as an order of magnitude, 
the outer discrepancy is less spectacular but
not less important. 

\section{The new density discrepancy from a new NFW halo phenomenology}

\begin{figure*}
   \centering
   \includegraphics[width=10.0cm]{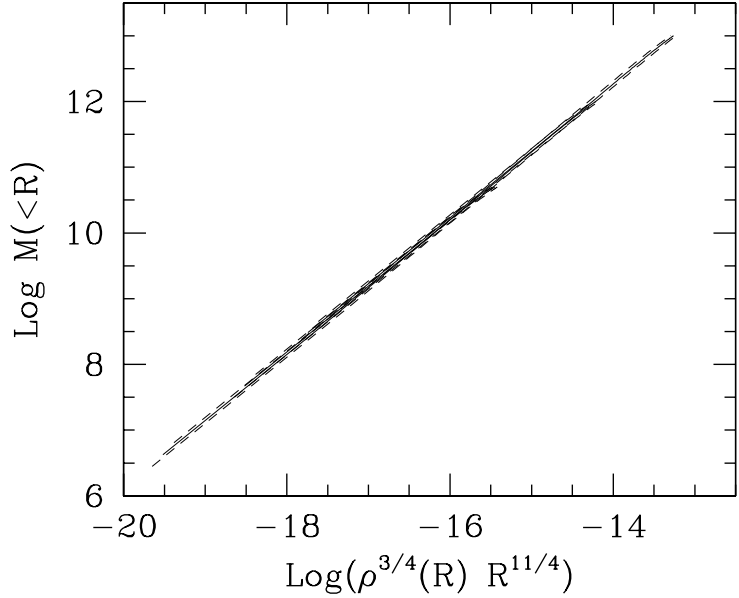}    
   \caption{
The $M-R-\rho$ relation: the (overlapping) 
solid lines refer to the $\Lambda$CDM haloes, for
3 different virial masses ($5 \times 10^{10}$ M$_{\odot}$, 
$1 \times 10^{12}$ M$_{\odot}$ and $1 \times 10^{13}$
M$_{\odot}$) and the 3 corresponding concentration parameters $c$
according to Wechsler et al. (2002) (18.7, 12.7 and 9.4, respectively).
The dashed lines correspond to the
$\pm$ 1-$\sigma$ uncertainty in $c$ 
taken from Wechsler et al. (2002).
Radii range from 0.001$r_{\rm vir}$ to $r_{\rm vir}$. 
M($<$R) is in M$_{\odot}$, 
$\rho$(R) in g cm$^{-3}$ and R in kpc.
}
\label{rela_all}
\end{figure*}

A main  property of the distribution of  NFW-CDM haloes is that   
$M_{\rm NFW}(R)= f(R, M_{\rm vir}, c(M_{\rm vir}))$, 
while $ \rho_{\rm NFW}(R)=h(R, M_{\rm vir}, c(M_{\rm vir}))$.
From this  
it follows that: $ M_{\rm NFW}(R) = {\cal G}(R, \rho_{\rm NFW}(R),M_{\rm vir})$, 
that is, at a fixed radius, 
the density and mass are
related, though in principle in a different way in galaxies of different $M_{\rm vir}$. 
The above provides us with a convenient way to present the structural properties of NFW
halos, according to which observational data can be compared with the theory
{\it without knowing the virial masses of the objects}. 
This is crucial because of the poor fitting performance of the NFW
fits to rotation curves often prevents even a rough estimate
of this quantity.
The relation we find
(hereafter called the $M-R-\rho$ relation)
is valid for any $M_{\rm vir}$; this can be easily seen in Fig. \ref{rela_all} by plotting it
for 3 different virial (total) masses ($5 \times 10^{10}$ M$_{\odot}$, 
$1 \times 10^{12}$ M$_{\odot}$ and $1 \times 10^{13}$
M$_{\odot}$) and concentration parameters (see eq. \ref{eq:wechsler2}), 
for radii ranging
from 0.001$r_{\rm vir}$ to $r_{\rm vir}$ the relation takes the
form (see Fig. \ref{rela_all}):

\begin{equation}
{\rm{log}}~ M_{\rm NFW} = \frac{3}{4} {\rm{log}}~
\rho_{\rm NFW}(R) +\frac{11}{4} {\rm{log}}~ R~+26.17 
\label{eq_rela}
\end{equation}

where $M_{\rm NFW}$ is in $M_{\odot}$, $R$ in kpc and $\rho_{\rm NFW}$ in 
g cm$^{-3}$.
The $M-R-\rho$ relation holds for more than 6 orders of magnitude
in mass, and it is valid, for real cases, 
for any galactic halo at any radius. 
The scatter in $c$, 3 orders of magnitude in radii and a factor 200 in $M_{\rm vir}$ 
introduce a negligible ($\lesssim$ 0.05 dex) scatter through 
the 6 orders of magnitude in $M_{\rm NFW}(R)$. 

The relation follows from the structural properties of NFW halos.
It can be derived mathematically, but its physical meaning is
reported to the origin of eqs. \ref{eq:rho_nfw} and \ref{eq:wechsler1}.
Let us write the ratio between the enclosed
mass $M_{\rm NFW}(<R)$ of an NFW halo at radius $r$ (where the density is
$\rho_{\rm NFW}(R)$) and a uniform sphere of radius $r$
and density $\rho_{\rm NFW}(R)$:

\begin{equation}
\frac{M_{\rm NFW}(<r)}{4/3 \pi r^3 \rho_{\rm NFW}(R)} = \frac{3 (1+cx)^2}{c^2 x^2}
\left({\rm ln}(1+cx)-\frac{cx}{1+cx}\right)
\end{equation}

On the other hand, from eq. \ref{eq_rela} we have: $\frac{M_{\rm NFW}(<r)}
{4/3 \pi r^3 \rho_{\rm NFW}(R)} \propto (\rho_{\rm NFW}(R) R)^{-1/4}
$. Then, recalling that $(\rho_{\rm NFW}(R) R)^{-1/4}=
\frac{(1+cx)^{1/2}}{(\rho_{\rm s} r_{\rm s})^{1/4}}$, one finds that
the $M-R-\rho$ relation is a way of expressing the approximation
(valid for $0<x<1$):

\begin{equation}
(1+cx)^{1/2} \propto A \frac{3 (1+cx)^2}{(c^2 x^2)}\left({\rm
    ln}(1+cx)-\frac{cx}{1+cx} \right)
\label{eq_approx}
\end{equation}

\begin{figure*}
\centering
\includegraphics[width=11cm]{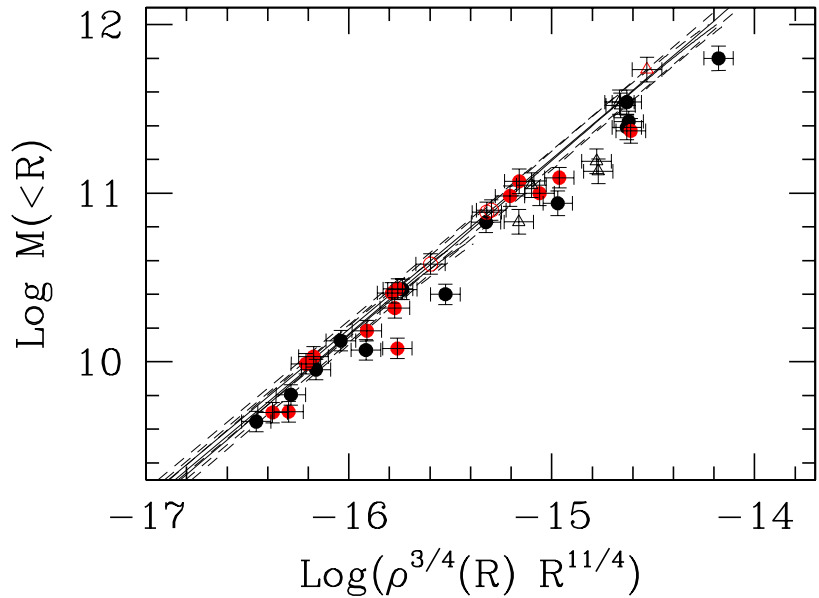}    
   \caption{
The same as Fig \ref{rela_all}, zoomed to the radial range
from 1\% to 20\% of the virial radius. 
Filled circles are the observational data of samples 1 and 2 at $R_{\rm f}$.
The three edge-on galaxies are denoted by empty circles. 
Empty triangles denote
the objects of sample 2 without published mass modelling. 
Red symbols are isolated halos and black symbols are subhalos, according
to our definition in the text. 
}
\label{rela}
\end{figure*}

We compared the predicted $M-R-\rho$ relation with the 
measurements of 37 galaxies.
The result (Fig. \ref{rela}) is that the observational points at $R_{\rm f}$
are systematically offset compared to the $M-R-\rho$ relation. 
The mean offset of $\sim$0.1 dex is solid (but note that larger offsets
are also observed); in fact, 
the error propagation analysis shows that a 3\% uncertainty on the measure of
the circular velocity and a 0.05 uncertainty on the logarithmic gradient
$\frac{d{\rm log}~V(R)}{d{\rm log}~R}$ yield uncertainties of the order
of 0.025 dex in $M$ and 0.06 dex in $\rho$.   
The uncertainties on the objects distances
are not included since they only
induce random errors. 
At $R_{\rm f}$, for a given enclosed mass $M(<R_{\rm f})$, 
the observed density $\rho_{\rm PI}(R_{\rm f})$
is higher than that of NFW haloes that match $M(<R_{\rm f})$.
The density discrepancy runs
up to a factor $\sim$3 and it
is significant in that the adopted procedure of matching the CDM mass to the
observed one is a conservative one.
The predicted halo profiles are based on a larger number of halos than
the present sample of 37 galaxies, so selection or systematic effects cannot
be completely excluded, even if the 37 galaxies span a large range of 
maximum velocities, Hubble types and environments.

We tested whether the environment has
an effect on our results: we distinguished between galaxies in
``isolated'' halos and galaxies in ``subhalos''. A straightforward definition
does not exist, so we defined as ``subhalos'' those galaxies which either
belong to a known group of galaxies or that have a larger companion within
30 $R_{25}$ and 400 km s$^{-1}$. The other galaxies were labelled as
"isolated". A more detailed investigation of the effect
of environment goes beyond the scope of our paper, especially because
of the heterogeneity of the data we consider here.
Qualitatively there
are no obvious differences between the different subsamples, 
even though the predicted difference
between the concentrations of halos and subhalos (Bullock et al. 2001)
cannot be excluded.


Edge-on galaxies may represent a potential problem, as the H$\alpha$ rotation
curves could be affected by extinction (Bosma et al. 1992) and the HI rotation curves might
suffer from projection effects unaccounted for and from the possible lack of gas on the line 
of nodes. 
In Fig. \ref{rela} we have plotted the three edge-on galaxies (i.e., with an inclination
larger than 85$^{\circ}$) of our sample with a triangle;
we realise that these galaxies do not lie 
in any peculiar region of the plane and they are not more discrepant than the other 
galaxies. The same holds for the galaxies from sample 2 for which the mass
model was obtained as described in the previous Section.

\begin{figure}
\centering
\includegraphics[width=8.5cm]{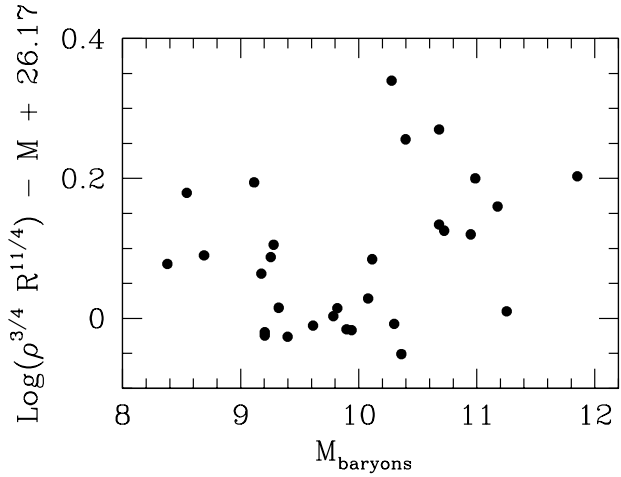}    
   \caption{
Distance from eq. \ref{eq_rela}
vs. baryonic mass of the galaxies with mass decompositions:
there is no clear correlation between the two quantities.
Units are the same as Fig. \ref{rela_all}.
}
\label{diff}
\end{figure}

Baryons are expected to affect the density distribution of dark matter,
but the effects are far from being clear: the best studied process is
adiabatic contraction (Blumenthal et al. 1986, Gnedin et 
al 2004, Sellwood \& McGaugh 2005),
which would make the halos more centrally concentrated,
but mechanisms with opposite effects have also been studied,
such as adiabatic expansion (Dutton et al. 2006) or
dynamical friction (Tonini, Lapi \& Salucci 2006).
Hence, in the $\Lambda$CDM halos considered here we did not take
the effect of baryons into account, since their effect on the 
distribution of dark matter is still poorly understood.
Also, baryons, when gas cools, are also expected to make dark matter 
halos more spherical (Kazantzidis et al. 2004). 
In Fig. \ref{diff} we plotted the baryonic mass of the galaxies of our 
sample with mass decompositions vs. their distance from the relation
in eq. \ref{eq_rela}: no clear correlation is observed, meaning that 
the discrepancy
we report in the present paper is not straightforwardly related to the
amount of baryons present in the galaxy, or that effects such as
uncertainties in the mass distribution create a large scatter
in Fig. \ref{diff}.
We also note that in the galaxies studied here $R_{\rm f} >> R_{\rm D}$,
i.e. $R_{\rm f}$ probes a region outside that most affected by
adiabatic contraction.

Warps and non-circular motions are also a potential concern
for the present analysis, since in a $\Lambda$CDM Universe
dark matter halos are expected to be triaxial, which would induce
non-circular motions in the gas (Hayashi \& Navarro 2006),
and gas moving through filaments 
(Dekel \& Birnboim 2006) will interact with galaxies,
triggering structures on disks like those warps
Nearly all the rotation curves collected in the present paper
were derived using the tilted-ring fitting of the velocity field,
which can account for warps but not for non-circular motions.
The exception is DDO 47, which was studied in detail by Gentile et al. (2005),
using the harmonic decomposition of the velocity field (Wong, Blitz \& Bosma 2004).
So we exclude that warps might have an effect on the paper conclusions,
while we cannot exlude the possibility of non-circular motions; however,
one can expect them to increase the observational scatter but not to
have a systematic effect.

The outer density discrepancy arises in a clear way by setting the NFW halo mass within
the last point equal to the observed one at $R_{\rm f}$; what happens if this
is valid at a different radius? If the radius is smaller, the density inversion
happens at even smaller radii, and the discrepancy at $R_{\rm f}$ is worse.
If the radius is larger, there might be no density inversion, but the 
cusp/core discrepancy (see however Section \ref{intro})
between $\Lambda$CDM predictions and observations would be increased and would
be present at any radius where baryons are present. 
The latter case is shown in Fig. \ref{flattening}, where at the last point
we imposed $\rho_{\rm h}(R_{\rm f})=\rho_{\rm NFW}(R_{\rm f})$ instead of
$M_{\rm h}(R_{\rm f})=M_{\rm NFW}(R_{\rm f})$: the inner discrepancy becomes worse.  
Even though the method of analysis is different, this
effect might be related to the results of Seigar et al. (2006), whose
NFW models (including possible adiabatic contraction) match the observed 
rotation curves of two galaxies in their outer
parts ($\sim$ 10 kpc), but overestimated the inner rotation curves.
One of the two cases shown by Seigar et al. is a barred galaxy
and its fitted concentration parameter is small compared to the averaged 
predicted value; the authors discuss possible evidence for the absence of  
adiabatic contraction or alternatively of some dynamical effects that  
compensates for adiabatic contraction.

\begin{figure*}
\centering
\includegraphics[width=11cm]{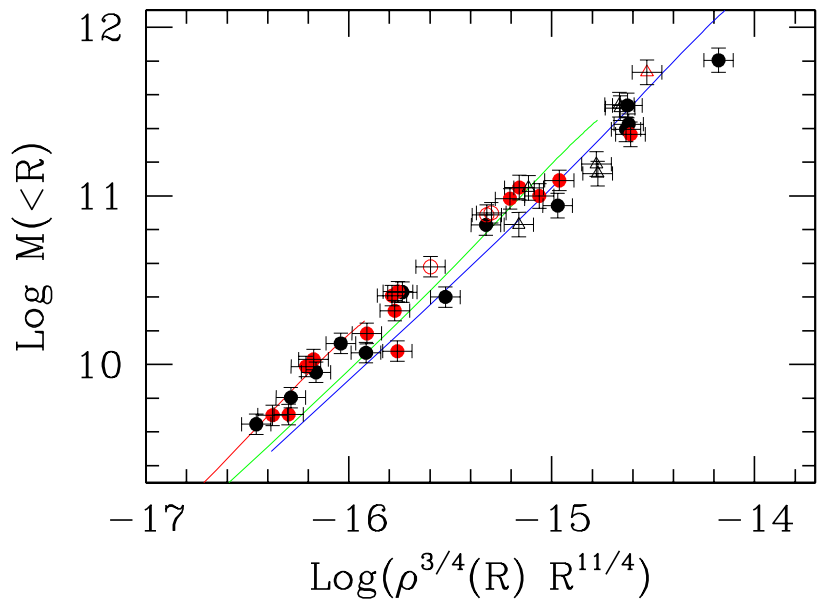}    
   \caption{
The same as Fig \ref{rela}, 
but showing that Burkert halos (Salucci \& Burkert 2000)
predict a mass dependent $M vs (\rho^a~R^b)$ relation
in good agreement with observations.
Symbols are the same as Fig \ref{rela}, and the 3 lines correspond to 
3 different virial masses ($5 \times 10^{10}$ M$_{\odot}$, 
$1 \times 10^{12}$ M$_{\odot}$ and $1 \times 10^{13}$
M$_{\odot}$, see text).
}
\label{rela_bur}
\end{figure*}

By assuming the Burkert halo instead of the NFW halo leads to the results
shown in Fig. \ref{rela_bur}. Here we consider the one-parameter 
family of halos defined by the Burkert halo and the empirical 
relation between the central density $\rho_0$ and the core radius $r_0$ 
found by Salucci \& Burkert (2000): $\rho_0=3 \times 10^{-24} 
(r_0/{\rm kpc})^{-2/3} {\rm g~ cm}^{-3}$. 
The virial masses were computed by integrating the density profile
until the mean density was $\Delta$ times $\rho_{\rm c}$, where 
$\Delta$ was derived following Bryan \& Norman (1998).
In this case (see Fig. \ref{rela_bur}) we find that halos with different masses
do not overlap (because for the chosen axes there is no such an approximation
as eq. \ref{eq_approx}), the agreement with the observations is much better
than in the case of the NFW halos. 
This means that the Burkert halo, known to fit the inner parts rotation 
curves better, is also a better representation of the observations at
the last radii probed by rotation curves ($\sim 5 - 35$\% of the virial
radius).

Let us point out that while the present work shows that a theory 
vs observations discrepancy extends well beyond the very inner 
regions of spirals, it leaves open the possibility that, for radii 
$r \gtrsim 0.2 - 0.3 r_{\rm vir} $,  the density of the DM around galaxies converges
to a NFW profile. This was hinted at by    
Prada et al. (2003) who
have investigated the very outer dark matter
profile by means of the kinematics of satellites around isolated galaxies.
However, they probe an outer region with respect to those considered
here and
in a very low resolution mode: their projected radii range from 20 to 350 kpc,
in 100 kpc bins.
Brainerd (2004a,b), using weak galaxy lensing in addition to the dynamics of
satellite galaxies, reach similar conclusions for scales $\gtrsim 50~h^{-1}$ kpc.
The results showed in the present paper hold for galaxies but the situation
on the scales of galaxy clusters might be different: indeed, Vikhlinin et al. (2006)
and Zappacosta et al. (2006) show agreement between X-ray data and 
$\Lambda$CDM mass profiles.

\section{Conclusions}

An accurate mass modelling of the external regions  in the case of   
a couple of test-case spirals  and a careful determination  of  the 
densities and enclosed masses of the dark matter  haloes at  the farthest radii  at which  
37 high quality rotation curves have been measured, has brought to the discovery 
of a new problem/discrepancy for the $\Lambda$CDM/NFW haloes.

In fact, in addition to the well-known evidence for which in the inner 
regions of galaxies ($R< 2 R_{\rm D}$) the DM haloes show a flattish density 
profile, with amplitudes up to  one order of magnitude lower than the $\Lambda$CDM  
predictions, at outer radii  ($R> 4  R_{\rm D}$)
the measured DM halo densities are found higher than the 
corresponding $\Lambda$CDM ones. This implies either that the shallow-steep
disagreement extends all over 1-2 times the galaxy optical radius 
(which would be the case if $\rho_{\rm NFW}(R_{\rm f}) = \rho(R_{\rm f})$)
or that
there is a complex data vs. theory disagreement.

While the statistical significance and the  level of the discrepancy 
must be investigated  with more and outer data, there is already an evidence for  
this discrepancy in most galaxies with high quality data. The DM halo 
density, known to have a core in the internal regions, does not seem to 
converge to the  NFW profile at 4-6 $R_{\rm D}$. 
This implies an issue for
$\Lambda$CDM that should be investigated
in future, when, due to improved observational techniques, the kinematic
information will be extended to the $\sim 100$ kpc scale (Gentile et al., in prep.).

This new discrepancy
provides additional information on the nature of the cusp/core issue:
self-interacting or annihilating dark matter proposed as the cause 
for the inner discrepancy may be in difficulties in that it will cause a
rapid convergence to the NFW profile in the luminous parts of galaxies
and beyond once a critical density 
value is reached. The discrepancy points to a scenario of modified CDM
profiles, to a global mass or angular momentum rearrangement
(e.g. Dutton et al. 2006, Tonini et al. 2006) that would 
remove dark matter from the innermost parts to
the radii probed by the outermost regions of rotation curves.

\begin{appendix}
\section{Persic \& Salucci (1990) method to derive the disk mass}

The total velocity profile is given by assuming the total centrifugal
equilibrium between the two components, the
baryons (only the disk is taken into account, as the bulge just affects
the centre of the rotation curve (RC), 
and the stellar halo mass and the HI disk are negligible for our
purposes) and the Dark Matter:
\begin{equation}
V^2(R)=V^2_{\rm D}(R)+V^2_{\rm H}(R)
\label{ecentrifugal}
\end{equation}
The halo and disk mass inside a given radius $R$ is
$M_{\rm H}(R)=G^{-1} \ V^2_{\rm H}(R) \ R$ and 
the disk contribution to the circular velocity is:
$V^2_{\rm D}(R)=G M_{\rm D} f/R$
respectively, where $f=1/2 \ (R/R_{\rm D})^2 \ (I_0K_0-I_1K_1)|_{R/2R_{\rm D}}$
is defined in terms of the modified Bessel functions $I_n$ and $K_n$. 
Let us now consider the moments of equation (\ref{ecentrifugal}), 
and let us define the logarithmic slopes of the total, disk and
halo velocity respectively: 
$
\nabla \equiv \frac{d \ log \ V(R)}{d \ log \ R} \Big\arrowvert_{R_{\rm opt}},
$
$
\nabla \equiv \frac{d \ log \ V_{\rm D}(R)}{d \ log \ R} \Big\arrowvert_{R_{\rm opt}}=-0.27,
$
$
\nabla_{\rm H} \equiv \frac{d \ log \ V_{\rm H}(R)}{d \ log \ R} \Big\arrowvert_{R_{\rm opt}}
$
where $R_{\rm opt}=3.2R_{\rm D}$. While $\nabla$ is observed, $\nabla_{\rm H}$ is unknown,
  and it is related to the Dark Matter distribution by:
$d(log \ M_{\rm H}(R))/d(log \ R)_{R_{\rm opt}} = 2\nabla_{\rm H} + 1$.
Since $M_{\rm D} \approx 1.1 V_{\rm D}^2(R) R/G$.
we can now write 
the disc-to-total mass ratio at the optical radius $R_{\rm opt}$:
\begin{equation}
\frac{M_{\rm D}(R_{\rm opt})}{M_{\rm total}(R_{\rm opt})} =\frac{\nabla_
{\rm H}-\nabla}{\nabla_{\rm H}+0.11\nabla+0.30}
\label{eratio}
\end{equation}
It is reasonable to assume that at the optical radius the halo RC is not
decreasing, and the DM density is not increasing, therefore $0 \leq
\nabla_{\rm H}
\leq 1$; in the maximum disk hypothesis, we can find the value of 
$\nabla_{\rm H}$
in the interval $[0,1]$ that maximises the ratio of equation (\ref{eratio}):
$\frac{d[M_{\rm D}(R_{\rm opt})/M_{\rm total}(R_{\rm opt})]}{d\nabla_{\rm H}}=0$
and in this case it is easy to see that 
$\nabla_{\rm H} = 1$.
From equation (\ref{eratio}) we then obtain:
\begin{equation}
M_{\rm D} \simeq 1.32 \ V^2_{\rm opt}R_{\rm opt} \ (0.77-0.83\nabla)
\label{eratio2}
\end{equation}
which allows us to obtain the halo velocity profile at any given point in its
luminous and dark contributions, by means of $M_{\rm D}$, $f$ 
and eq. \ref{ecentrifugal}. 

Notice that this minimum halo assumption used to derive the disk mass might
not be completely correct and it might enhance the evidence for an
inner core, but it is perfectly legitimate in the present procedure in that
in any case it just (slightly) under-estimates the DM
density at large radii, bringing even more support to our claim.

In any case the disk mass, obtained by means of eq. \ref{eratio2},
plays a very minor role in the results
of this work, 
in fact at large distances 
the mass and density of the dark matter at any outermost radius
(denoted as $R_{\rm f}$) do not depend significantly on this quantity. 
We have,
supposing to deal with a system in virial equilibrium:
\begin{equation}
\rho(R_{\rm f})= \frac{V^2(R_{\rm f})}{4\pi GR_{\rm f}^2} \ \left( 1 +
2\frac{dlogV(R)}{dlogR}\Big\arrowvert_{R_{\rm f}} \right) + \frac{M_{\rm
    D}}{R_{\rm D}^3} f(R_{\rm f}/R_{\rm D})
\label{erhofinale}
\end{equation}
(see Fall \& Efstathiou 1980). We realise that, at most, the second term
is a correction of $30\%$. So 
errors (even quite large) in the estimate of $M_{\rm D}$ little propagate in the
estimate of $\rho(R_{\rm f})$. 

\end{appendix}


\begin{thebibliography}{}

\bibitem[\protect\citeauthoryear{Barnes et al.}{2004}]{barnes}
Barnes, E.I., Sellwood, J.A., Kosowsky, A., 2004, AJ, 128, 2724

\bibitem[\protect\citeauthoryear{Bell et al.}{2003}]{B:03} Bell, E.F.,
  McIntosh, D.H., Katz, N., Weinberg, M.D., 2003, ApJS, 149, 289

\bibitem[\protect\citeauthoryear{Blais-Ouellette et al.}{2001}]{blais}
  Blais-Ouellette, S., Amram, P., Carignan, C., 2001, AJ, 1952, 1964

\bibitem[\protect\citeauthoryear{Blais-Ouellette et al.}{2004}]{blais2}
Blais-Ouellette, S., Amram, P., Carignan, C., Swaters, R., 2004, A\&A, 420, 147

\bibitem[\protect\citeauthoryear{Brainerd}{2004a}]{brainerda}
Brainerd, T. G., 2004a, preprint (astro-ph/0409381)

\bibitem[\protect\citeauthoryear{Brainerd}{2004b}]{brainerdb}
Brainerd, T. G., 2004b, AIP Conf.~Proc.~743: The New Cosmology: Conference 
on Strings and Cosmology, 743, 129 

\bibitem[Blumenthal et al.(1986)]{1986ApJ...301...27B} Blumenthal, G.~R., 
Faber, S.~M., Flores, R., \& Primack, J.~R., 1986, ApJ, 301, 27 

\bibitem[\protect\citeauthoryear{Borriello \& Salucci}{2001}]{borriello_salucci}
Borriello, A., Salucci, P., 2001, MNRAS, 323, 285

\bibitem[\protect\citeauthoryear{Bosma et al.}{1992}]{bosma}
Bosma, A., Byun, Y., Freeman, K. C., Athanassoula, E., 1992, ApJ, 400, L21

\bibitem[\protect\citeauthoryear{Bottema}{1999}]{bottema}
Bottema, R., 1999, A\&A, 348, 77

\bibitem[\protect\citeauthoryear{Bryan \& Norman}{1998}]{bryan}
Bryan, G.-L., Norman, M.-L., 1998, ApJ, 495, 80

\bibitem[\protect\citeauthoryear{Burkert}{1995}]{burkert}
Burkert, A., 1995, ApJ, 447, L25

\bibitem[\protect\citeauthoryear{Carignan et al.}{1990}]{carignan}
Carignan, C., Charbonneau, P., Boulanger, F., Viallefond, F., 1990, 
A\&A, 234, 43

\bibitem[\protect\citeauthoryear{de Blok \& Bosma}{2002}]{deblok1}
de Blok, W. J. G., Bosma, A., 2002, A\&A, 385, 816

\bibitem[\protect\citeauthoryear{de Blok et al.}{2003}]{deblok2}
de Blok, W.~J.~G., Bosma, A., McGaugh, S.\ 2003, MNRAS, 340, 657 

\bibitem[\protect\citeauthoryear{de Blok}{2005}]{deblok3}
de Blok, W.~J.~G.\ 2005, ApJ, 634, 227 

\bibitem[Dekel \& Birnboim(2006)]{2006MNRAS.368....2D} Dekel, A.,  
Birnboim, Y., 2006, MNRAS, 368, 2 

\bibitem[\protect\citeauthoryear{Donato et al.}{2004}]{donato}
Donato, F., Gentile, G., Salucci, P., 2004, MNRAS, 353, L17

\bibitem[\protect\citeauthoryear{Dutton et al.}{2005}]{dutton}
Dutton, A. A., Courteau, S., de Jong, R., Carignan, C., 2005, 619, 218

\bibitem[\protect\citeauthoryear{Dutton et al.}{2006}]{dutton06}
Dutton, A. A., van den Bosch, F. C., Dekel, A., Courteau, S., 2006, preprint
(astro-ph/0604553)

\bibitem[\protect\citeauthoryear{Gentile et al.}{2004}]{gentile}
Gentile, G., Salucci, P., Klein, U., Vergani, D., Kalberla, P.,
2004, MNRAS, 351, 903

\bibitem[\protect\citeauthoryear{Gentile et al.}{2005}]{gentile2}
Gentile, G., Burkert, A., Salucci, P., Klein, U., Walter, F.,
2005, ApJ, 634, L145

\bibitem[\protect\citeauthoryear{Gentile et al.}{2006}]{gentile3}
Gentile, G., Salucci, P., Klein, U., Granato, G. L., 
2006, MNRAS in press, astro-ph/0611355

\bibitem[\protect\citeauthoryear{Giraud}{1998}]{giraud}
Giraud, E., 1998, AJ, 116, 2177

\bibitem[Gnedin et al.(2004)]{2004ApJ...616...16G} Gnedin, O.~Y., Kravtsov, 
A.~V., Klypin, A.~A., \& Nagai, D., 2004, ApJ, 616, 16 

\bibitem[\protect\citeauthoryear{Gnedin}{2006}]{gnedin}
Gnedin, O. Y., Weinberg, D. H., Pizagno, J., Prada, F., Rix, H.-W., 
2006, preprint (astro-ph/0607394)

\bibitem[\protect\citeauthoryear{Hayashi \& Navarro}{2006}]{hayashi}
Hayashi, E., Navarro, J. F., 2006, MNRAS, in press, astro-ph/0608376

\bibitem[Kazantzidis et al.(2004)]{2004ApJ...611L..73K} Kazantzidis, S., 
Kravtsov, A.~V., Zentner, A.~R., Allgood, B., Nagai, D., Moore, B., 
2004, ApJ, 611, L73 

\bibitem[\protect\citeauthoryear{McGaugh et al.}{2006}]{M:06}
McGaugh, S.~S., de Blok, W.~J.~G., Schombert, J.~M., Kuzio de Naray, R., \& Kim, J.~H.\ 2006, 
ApJ, in press (astro-ph/0612410) 

\bibitem[\protect\citeauthoryear{Navarro, Frenk and White}{1996}]{NFW:96}
  Navarro, J.F., Frenk, C.S., White, S.D.M., 1996, ApJ, 462, 563

\bibitem[\protect\citeauthoryear{Noordermeer et al.}{2004}]{N:04}
Noordermeer, E., van der Hulst, T., Swaters, R., 2004, Proceedings of "Baryons
in Dark Matter Halos". Novigrad, Croatia, 5-9 Oct 2004. Eds: R. Dettmar, U. Klein, P. Salucci. Published by SISSA, Proceedings of Science, http://pos.sissa.it, p. 68.

\bibitem[\protect\citeauthoryear{Persic and Salucci}{1990}]{PS:90}
Persic, M., Salucci, P., 1990, MNRAS, 247, 349

\bibitem[\protect\citeauthoryear{Prada et al.}{2003}]{Pr:03} Prada, F., Vitvitska, M., Klypin, A., Holtzman, J. A., Schlegel, D. J., Grebel, E. K., Rix, H.-W., Brinkmann, J., McKay, T. A., Csabai, I., 2003, ApJ, 598, 260

\bibitem[\protect\citeauthoryear{Salucci \& Burkert}{2000}]{Sal:00} Salucci, P., Burkert, A., 2000, ApJ, 537, L9

\bibitem[\protect\citeauthoryear{Salucci et al.}{2003}]{Sal:03} Salucci, P., Walter, F., Borriello, A., 2003, A\&A, 409, 53

\bibitem[\protect\citeauthoryear{Seigar et al.}{2006}]{Sei:06}
Seigar, M.~S., Bullock, J.~S., Barth, A.~J., \& Ho, L.~C.\ 2006, ApJ, 645, 1012 

\bibitem[Sellwood \& McGaugh(2005)]{2005ApJ...634...70S} Sellwood, J.~A., 
McGaugh, S.~S.\, 2005, ApJ, 634, 70 

\bibitem[\protect\citeauthoryear{Simon et al.}{2005}]{Si:05} Simon, J.D., Bolatto, A.D., Leroy, A., Blitz, L., Gates, E.L., 2005, ApJ, 621, 757

\bibitem[\protect\citeauthoryear{Sofue et al.}{1999}]{So:99}
Sofue, Y., Tutui, Y., Honma, M., Tomita, A., Takamiya, T., Koda, J., Takeda,
Y., 1999, ApJ, 523, 136

\bibitem[\protect\citeauthoryear{Spergel et al.}{2006}]{Sp:06}
Spergel, D. N., et al., 2006, preprint (astro-ph/0603449)

\bibitem[\protect\citeauthoryear{Swaters et al.}{2003}]{S:03}
 Swaters, R.A., Madore, B.F., van den Bosch, F.C., Balcells, M., 2003, ApJ, 583, 732

\bibitem[Tonini et al.(2006)]{2006ApJ...649..591T} Tonini, C., Lapi, A.,  
Salucci, P., 2006, ApJ, 649, 591

\bibitem[\protect\citeauthoryear{Valenzuela et al.}{2007}]{Va:07}
Valenzuela, O., Rhee, G., Klypin, A., Governato, F., Stinson, G., Quinn, T., Wadsley, 
J.\ 2007, ApJ, in press, astro-ph/0509644

\bibitem[\protect\citeauthoryear{van Albada et al.}{1985}]{vA:85}
van Albada, T.S., Bahcall, J.N., Begeman, K., Sancisi, R., 1985, ApJ, 295, 305

\bibitem[\protect\citeauthoryear{van den Bosch et al.}{2000}]{vdB:00} van den Bosch, F.C., Robertson, B.E., Dalcanton, J., de Blok, W.J.G., 2000, AJ, 119, 1579

\bibitem[\protect\citeauthoryear{van Zee and Bryant}{1999}]{vZ:99}
van Zee, L., Bryant, J., 1999, AJ, 118, 2172

\bibitem[\protect\citeauthoryear{Verheijen and Sancisi}{2001}]{VS:01}
Verheijen, M. A. W., Sancisi, R., 2001, A\&A, 370, 765

\bibitem[\protect\citeauthoryear{Walsh et al.}{1997}]{W:97}
Walsh, W., Staveley-Smith, L., Oosterloo, T., 1997, AJ, 113, 1591

\bibitem[\protect\citeauthoryear{Wechsler et al.}{2002}]{We:02} Wechsler, R.H., Bullock, J.S., Primack, J.R., Kravtsov, A.V., Dekel, A., 2002, ApJ, 568, 52

\bibitem[\protect\citeauthoryear{Weiner et al.}{2002}]{We:01}	
Weiner, Benjamin J., Williams, T. B., van Gorkom, J. H., Sellwood, J. A.,
2001, ApJ, 546, 916

\bibitem[\protect\citeauthoryear{Weldrake, de Blok \& Walter}{Weldrake et al.}{2003}]{W:03}
 Weldrake, D.T.F., de Blok, W.J.G., Walter, F., 2003, MNRAS, 340, 12

\bibitem[\protect\citeauthoryear{Wong, Blitz \& Bosma}{2004}]{Wo:04}
Wong, T., Blitz, L., Bosma, A., 2004, ApJ, 605, 183	

\end{thebibliography}
\end{document}